\documentclass[twocolumn,aps,amsmath,prd,amssymb,eqsecnum,nofootinbib]{revtex4-1}

\usepackage[colorlinks=true,citecolor=blue]{hyperref}
\usepackage{graphicx}
\usepackage{multirow}
\usepackage[normalem]{ulem}

\begin{document}

\title{Propagation of Test Particles and Scalar Fields on a Class of Wormhole Space-Times}

\author{Peter Taylor}
\email{petertaylor@astro.cornell.edu}
\affiliation{Center for Radiophysics and Space Research, Cornell University, Ithaca, NY 14853, USA\\
 and\\
School of Mathematical Sciences and Complex \& Adaptive Systems Laboratory, University College Dublin, UCD, Belfield, Dublin 4, Ireland}

\date{\today}
\begin{abstract}
In this paper, we consider the problem of test particles and test scalar fields propagating on the background of a class of wormhole space-times. For test particles, we solve for arbitrary causal geodesics in terms of integrals which are solved numerically. These integrals are parametrized by the radius and shape of the wormhole throat as well as the initial conditions of the geodesic trajectory. In terms of these parameters, we compute the conditions for the geodesic to traverse the wormhole, to be reflected by the wormhole's potential or to be captured on an unstable bound orbit at the wormhole's throat. These causal geodesics are visualized by embedding plots in Euclidean space in cylindrical coordinates. For massless test scalar fields, we compute transmission coefficients and quasi-normal modes for arbitrary coupling of the field to the background geometry in the WKB approximation. We show that solutions of the scalar wave equation on this class of wormholes are stable only for certain values of the coupling constant. This analysis is interesting since recent computations of self-interactions of a static scalar field in wormhole space-times reveal some anomalous dependence on the coupling constant such as the existence of an infinite discrete set of poles. We show that this pathological behavior of the self-field is an artifact of computing the interaction for values of the coupling constant that do not lie in the domain of stability. \end{abstract}
\maketitle

\section{Introduction}
Wormholes are topological bridges connecting distant regions of the universe or in the multiverse scenario the wormhole may be a bridge between two different universes. Interest in wormhole space-times dates back to 1916 \cite{Flamm}, pre-dating interest in black hole space-times. Its modern popularity is owed primarily to the work of Morris and Thorne \cite{MorrisThorne} who investigated the idea of using so-called ``traversable wormholes'' as a means for time travel, albeit in the context of a novel pedagogical tool for teaching General Relativity. The distinction between traversable and non-traversable wormholes is that in the former case, the space-time is assumed to have no horizons and the gravitational tidal forces assumed to be bearable by a human traveller. Morris and Thorne showed that in the context of classical relativity, such space-times require a stress-energy tensor that violates the null energy condition, that is they require the existence of \textit{exotic} matter, for example ghost scalar fields or phantom energy \cite{ArmendarizPicon, Sushkov, LoboPhantomWH, GonzalezPhantomWH}. As such, wormholes were initially ruled out as objects of astrophysical relevance. However, violations of energy conditions is the \textit{status quo} in quantum field theories and this led to a surge of interest in wormhole solutions in semi-classical gravity (see \cite{VisserWormhole, Lobo} for comprehensive reviews). Moreover, Barcel\`o and Visser have shown \cite{BarceloVisser} that there are a number of apparently innocuous non-minimally coupled scalar field configurations that lie within the domain of current experimental constraints which violate the classical energy conditions and lead to traversable wormhole solutions. On the other hand, if one regards the classical energy conditions to be fundamental, there still exists traversable wormhole solutions in alternative theories of gravity \cite{Bertolami, GarciaLobo, LoboOliveiraBD, LoboOliveirafR, LoboWeylWH, HarkoLobo, Hohmann} and in higher dimensional theories \cite{LoboBraneWH, CorreaOliva, BronnikovKim} that obey the energy conditions near the wormhole throat. In fact, in Einstein-Gauss-Bonnet gravity in higher odd dimensions, one may even obtain static wormhole space-times that are vacuum solutions \cite{DottiOliva}. That wormhole solutions arise in a number of different physically reasonable higher dimensional and modified gravity configurations has given new impetus to their study, where the focus in the literature has been on their stability rather than their existence. For example, it is known that if Einstein's theory is coupled to a ghost scalar field via the action
\begin{align}
S[g,\varphi]=\int (\tfrac{1}{16\pi G}R+\tfrac{1}{2}g^{\mu\nu} \nabla_{\mu}\varphi\nabla_{\nu}\varphi)\sqrt{-g}\,d^{4}x,
\end{align}
then the theory admits static, spherically symmetric traversable wormholes which are unstable \cite{GonzalezI} and may collapse to a Schwarzschild black hole or expand indefinitely \cite{ShinkaiHayward, GonzalezII}. Traversable wormholes may also be supported by less exotic sources than ghost fields, for example, non-minimally coupled scalar fields but again these have been shown to be linearly unstable \cite{BronnikovGrinyok}. The issue may be circumvented by assuming that the wormhole throat is supported by a phantom thin shell \cite{LoboStableWH} or by adopting a modified theory of gravity \cite{KantiStableWH}. Very recently, the first stable wormhole in general relativity without invoking phantom thin shells has been reported \cite{BronnikovStableWH}. Hence, although wormholes in GR remain physically speculative, some of the obstacles to their viability in gravitational physics are slowly being eroded, particularly if one takes the modified gravity paradigms seriously.

In this article, we wish to characterize the propagation of test particles and scalar fields on a class of wormhole space-times. There is a long history of similar work in the literature and the key goal of this work is to extend in several directions the existing catalogue. In Sec.~\ref{sec:geodesics}, we consider causal geodesics in a class of wormhole space-times parametrized by the size and curvature of the wormhole throat. The geodesics in the Ellis wormhole were solved in terms of Elliptic integrals \cite{Muller} which is a specific case of the class of wormhole under consideration here. In general, the solutions cannot be given in terms of known functions but must be solved numerically. Nevertheless, the fate of a causal geodesic, i.e., whether the geodesic is reflected, captured on an unstable null orbit or traverses the wormhole, can be characterized in terms of the parameters of the wormhole and the initial conditions. The fact that there always exists an unstable null geodesic for the class of wormholes considered in this article is of interest for test scalar fields since it is well known that such orbits are related to the existence of a quasi-normal mode spectrum \cite{FerrariMashhoon2,Mashhoon}. Indeed this was one of the motivations for studying geodesics in this class of wormhole space-times. Finally in Sec.~\ref{sec:geodesics}, we show that by embedding the constant time, equatorial slices of the space-time in Euclidean space in cylindrical coordinates, a natural visualization of all causal geodesics is immediately amenable.

In Sec.~\ref{sec:scalar}, we turn our attention to the propagation of scalar fields on the background geometry. Notwithstanding the aforementioned issues of wormhole stability, one can still consider the stability of a scalar test field propagating on the background wormhole space-time, i.e., we ignore the back-reaction of the scalar on the wormhole geometry. We shall show that the stability of the solutions to the scalar wave equation is sensitive to the coupling of the scalar to the background geometry. When restricted to the stable domain of solutions, the effective potential is positive definite with a single peak which asymptotes to zero at both spatial infinities. Hence, the scattering problem is analogous to the black hole case and the methods developed for scattering off black holes are immediately applicable. We adopt the WKB method to fourth order to compute transmission coefficients and quasi-normal modes, extending the work of Ref.~\cite{KonoplyaZhidenko} to arbitrary coupling. To lowest order in the geometric optics limit, we retrieve the result of Ref~\cite{KonoplyaZhidenko}, but the next to leading order contains a dependence on the coupling.

Finally, we reconsider the calculation of the static scalar self-force in the Ellis wormhole space-time derived in \cite{Taylor}. We show that the restriction of the coupling constant to the domain of stability removes both the pathological behavior and the spurious vanishing of the self-force. Hence the static self-force is a smooth function that vanishes only at the conformal coupling value in three dimensions. The anomalous behavior of the self-field of a static scalar in wormholes space-times was first noted by Bezerra and Khusnutdinov \cite{Khusnutdinov2}. We argue that this behavior is generically an artifact of computing the self-field for values of the coupling constant outside the domain of stability. We illustrate this point in a simple wormhole where both the quasi-normal modes

\section{The Wormhole Geometry and Geodesics}
\label{sec:geodesics}
We begin with the Morris-Thorne \textit{ansatz} for a static wormhole metric in spherical polar coordinates
\begin{align}
\label{eq:metric}
ds^{2}=-e^{2\Lambda(r)}dt^{2}+(1-b(r)/r)^{-1}dr^{2}+r^{2}d\Omega^{2}
\end{align}
where  $\Lambda(r)$ is the lapse function or red-shift function, $b(r)$ is the shape function which determines the throat profile and $d\Omega^{2}$ is the metric on the two-sphere $\mathbb{S}^{2}$. We note that in the ultra-static case, $\Lambda(r)=0$, there is no gravitational acceleration in this frame, i.e., a particle dropped from rest remains at rest. In order for the space-time to be free from horizons, we require $\Lambda(r)$ to be everywhere finite while asymptotic flatness implies $\Lambda(r)\to 0$ and $b(r)/r\to 0$ as $r\to\infty$. The profile function $b(r)$ determines the shape of the wormhole throat, which can be visualized by taking the constant time slices in the equatorial plane of the wormhole metric and embedding in Euclidean space in cylindrical coordinates, where the embedding function is given by
\begin{align}
\frac{dz}{dr}=\pm\Big(\frac{r}{b(r)}-1\Big)^{-1/2}.
\end{align}
The coordinate $r$ runs from spatial infinity $r=\infty$ down to its minimum value $r_{\textrm{min}}=b(r_{\textrm{min}})\equiv b_{0}$ which is the location of the wormhole throat whence the tangent to the embedding function is vertical. Requiring the throat to be connecting two asymptotically flat regions implies the embedding function flares out in the vicinity of the throat, this is the so-called \textit{flaring-out} condition. Mathematically speaking, the inverse of the embedding function $r(z)$ must be concave up in the vicinity of the throat, i.e.,
\begin{align}
\label{eq:flareoutcondition}
 \frac{d^{2}r}{dz^{2}}=\frac{b(r)-r\,b'(r)}{2b(r)^{2}}>0\qquad \textrm{at or near the throat.}
 \end{align}
It is useful to introduce the proper radial distance, $\rho$, in terms of which the line-element is
\begin{align}
ds^{2}=-e^{2\Lambda(\rho)}dt^{2}+d \rho^{2}+r^{2}(\rho)d\Omega^{2}
\end{align}
where $\rho$ is related to $r$ by
\begin{align}
\label{eq:properradius}
\rho=\pm\int_{b_{0}}^{r}\Big(1-\frac{b(r)}{r}\Big)^{-1/2}dr.
\end{align}
We note that that there are two spatial infinities $\rho\to\pm\infty$ corresponding to $r\to\infty$ and the throat is located at $\rho=0$. 

The geodesic equations in the equatorial plane are given by
\begin{align}
\dot{t}&=E\,e^{-2\Lambda(r)}\nonumber\\
\dot{\phi}&=L/r^{2}\nonumber\\
\dot{r}^{2}&=\Big(1-\frac{b(r)}{r}\Big)\Big(E^{2}e^{-2\Lambda(r)}-\frac{L^{2}}{r^{2}}+\varepsilon\Big),
\end{align}
where $E$ is the conserved energy, $L$ the conserved momentum and $\varepsilon=-1,0$ for time-like and null geodesics, respectively. Differentiation is with respect to proper time for time-like geodesics and an affine parameter for null geodesics. If we restrict ourselves to the class of ultra-static wormholes $\Lambda=0$, then the geodesic equations in terms of the proper radial distance take the simple form
\begin{align}
\label{eq:geodesics}
\dot{t}=E,\qquad \dot{\phi}=L/r^{2}(\rho),\qquad
\Big(\frac{d \rho}{ds}\Big)^{2}=E^{2}-V(L,\rho),
\end{align}
where the potential is given by
\begin{align}
\label{eq:potential}
V(L,\rho)=\frac{L^{2}}{r(\rho)^{2}}-\varepsilon,
\end{align}
which is everywhere positive (for $L\ne 0$) and asymptotes to zero as $\rho\to\pm\infty$. A generic feature of this potential is that it possesses a global maximum at the throat, a fact which is readily verified by
\begin{align}
 \frac{dV}{d\rho}\Big|_{\rho=0}=0,\quad \frac{d^{2}V}{d\rho^{2}}\Big|_{\rho=0}=-\frac{L^{2}}{r^{4}}\Big(\frac{b(r)}{r}-b'(r)\Big)\Big|_{\rho=0}<0.
\end{align}
The inequality above is a result of the flaring out condition (\ref{eq:flareoutcondition}). If we assume that $r(\rho)$ is everywhere concave up, then $\rho=0$ is the only turning point in the potential and it delimits those geodesics coming in from spatial infinity $\rho=\infty$ which are reflected back to $\rho=\infty$ from those that pass through the wormhole throat and reach the other spatial infinity $\rho=-\infty$, with the limiting case at the peak of the potential corresponding to a bound orbit. This is clearly an unstable orbit since it occurs at the maximum of the potential. For $r(\rho)$ not necessarily globally concave up, we can have stable circular geodesics and potential wells admitting oscillatory motion between the turning points \cite{SarbachZannias}. Nevertheless, regardless of what form $r(\rho)$ takes, we always have an unstable circular geodesic at the throat. It is well established that the existence of an unstable null circular geodesic is intimately connected to the existence of a quasi-normal mode spectrum (see, e.g., \cite{FerrariMashhoon2}), which we compute in Sec.~\ref{sec:qnm}.

\begin{figure}
\centering
\includegraphics[width=8.3cm]{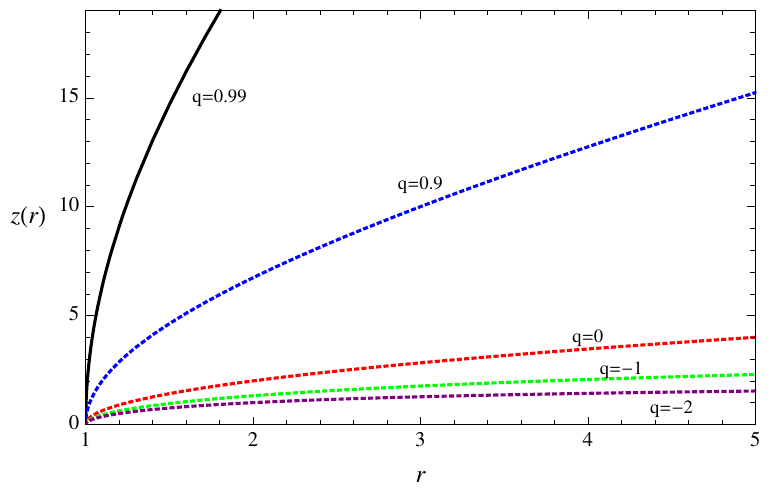}
\caption{{Plot of the embedding function $z(r)$ for various values of the shape exponent $q$. The throat width has been set to $b_{0}=1$.} }
\label{fig:embedding}
\end{figure}

\begin{figure}
\centering
\includegraphics[width=8.3cm]{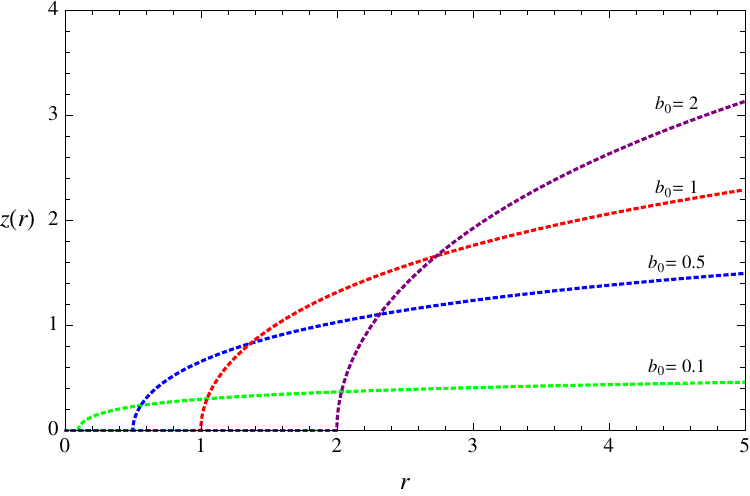}
\caption{{Plot of the embedding function $z(r)$ for various values of the throat radius $b_{0}$. The shape exponent has been set to $q=-1$, i.e., the Ellis wormhole. As the throat radius shrinks, the handle between the upper and lower universe becomes arbitrarily short. In the limit $b_{0}\to 0$, we retrieve two copies of Minkowski space-time.} }
\label{fig:embeddingthroatsize}
\end{figure}

In order to give a more concrete and quantitative analysis, we will henceforth restrict ourselves to a particular class of ultra-static wormholes defined by the shape function
\begin{align}
\label{eq:shapefn}
b(r)=b_{0}^{1-q}r^{q},\quad q<1.
\end{align}
The exponent $q$ we will refer to as the \textit{shape exponent}. The Ellis wormhole \cite{Ellis} corresponds to taking $q=-1$ and arbitrary geodesics in this space-time were comprehensively studied in Ref.~\cite{Muller}. For this particular class of power-law shape functions, the embedding function may be given explicitly as
\begin{align}
\label{eq:embedding}
z(r)=i\,r\,\,{}_{2}F_{1}\big(\tfrac{1}{2},\tfrac{1}{1-q},1+\tfrac{1}{1-q},(r/b_{0})^{1-q}\big)\nonumber\\
-i\,b_{0}\sqrt{\pi}\frac{\Gamma(1+\tfrac{1}{1-q})}{\Gamma(\tfrac{1}{2}+\tfrac{1}{1-q})},
\end{align}
where ${}_{2}F_{1}(\alpha,\beta,\gamma,t)$ is the Hypergeometric function \cite{GradRiz}. The embedding function for various values of the shape exponent and throat radius are plotted in Figs.~\ref{fig:embedding}-\ref{fig:embeddingthroatsize}.

\begin{figure}
\centering
\includegraphics[width=8.2cm]{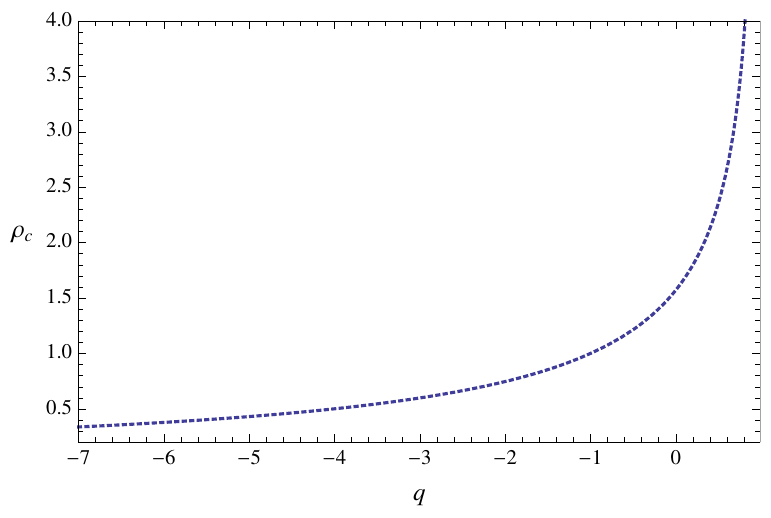}
\caption{{Plot of the radius of convergence $\rho_{\textrm{c}}$ as a function of the shape exponent $q$. The throat radius has been set to $b_{0}=1$.} }
\label{fig:convergence}
\end{figure}

Similarly, the proper radial distance defined by Eq.~(\ref{eq:properradius}) can be obtained explicitly in terms of Hypergeometric functions:
\begin{align}
\label{eq:rho}
\rho(r)=\pm \frac{2 b_{0}i}{3-q}\Big({}_{2}F_{1}\Big(\frac{1}{2},\frac{3-q}{2(1-q)},\frac{5-3q}{2(1-q)},\Big(\frac{r}{b_{0}}\Big)^{1-q}\Big)\nonumber\\
\times \Big(\sqrt{\frac{r}{b_{0}}}\Big)^{3-q}-\sqrt{\pi}\frac{\Gamma(\tfrac{5-3q}{2(1-q)})}{\Gamma(\tfrac{2-q}{1-q})}\Big).
\end{align}
Unfortunately, an exact explicit representation for the potential $V(L,\rho)$ as a function of $\rho$ is intractable since Eq.~(\ref{eq:rho}) cannot in general be inverted. However, the positive and negative branch of $\rho(r)$ are monotonic functions and can be numerically inverted using standard techniques. Alternatively, one can obtain a Taylor expansion for $V(L,\rho)$ about the throat and match to asymptotic series for large proper radii. For the Taylor expansion, all derivatives of $r(\rho)$ evaluated at the throat can be obtained by repeated differentiation of
\begin{align}
\frac{d r}{d\rho}=\pm\Big(1-\frac{b_{0}^{1-q}}{r^{1-q}}\Big)^{1/2}.
\end{align}
Since $r$ (and hence $V(L,\rho)$) is a symmetric function of $\rho$, all odd order derivatives of $V(L,\rho)$ vanish at the throat, and hence the Taylor expansion yields
\begin{align}
V(L,\rho)=\sum_{k=0}^{\infty}\frac{1}{2k!}V^{(2k)}(L,0)\rho^{2k}
\end{align}
where the coefficients up to $8^{\textrm{th}}$ order are
\begin{align}
V^{(0)}(L,0)&=\frac{L^{2}}{b_{0}^{2}}-\varepsilon\nonumber\\
V^{(2)}(L,0)&=\frac{L^{2}(q-1)}{2 b_{0}^{4}}\nonumber\\
V^{(4)}(L,0)&=-\frac{L^{2}(q-1)^{2}}{48 b_{0}^{6}}(q-11)\nonumber\\
V^{(6)}(L,0)&=\frac{L^{2}(q-1)^3}{720 b_{0}^{8}}(q^{2}-16 q+73)\nonumber\\
V^{(8)}(L,0)&=-\frac{L^{2}(q-1)^4}{161280 b_{0}^{10}}\left(17 q^{3}-354 q^{2}+2613 q-7096\right).
\end{align}
The radius of convergence of this series is the distance from the throat to the nearest singularity in the complex $\rho$-plane which occurs whenever $r=0$, which from Eq.~(\ref{eq:rho}) is given by
\begin{align}
|\rho|<\frac{2\,b_{0}\sqrt{\pi}}{(3-q)} \frac{\Gamma\left(\tfrac{5-3q}{2(1-q)}\right)}{\Gamma\left(\tfrac{2-q}{1-q}\right)}.
\end{align}
If the radius of convergence is sufficiently large, we can match to an asymptotic series for large $\rho$, otherwise we can patch with other Taylor expansions with overlapping convergence regions. A plot of this convergence radius as a function of the shape exponent is given in Fig.~\ref{fig:convergence}. It is a monotonically increasing function of the shape exponent, where we can have an arbitrarily large radius of convergence by taking $q$ to be sufficiently close to unity whereas the radius of convergence becomes arbitrarily small as $q$ becomes increasingly negative. For this reason, it is best to obtain $r(\rho)$ by numerically inverting (\ref{eq:rho}), nevertheless, for $q$ sufficiently close to unity, the Taylor expansion method yields very accurate plots for the potential.

We will assume that the geodesics are non-radial and non-circular since these cases follow trivially by setting $\dot{\phi}=0$ and $\dot{r}=0$ respectively in the equations (\ref{eq:geodesics}). In order to solve the geodesic equations requires specifying initial conditions. Following Ref.~\cite{Muller}, we wish to characterize all causal geodesics in terms of an initial position and direction with respect to a local reference frame of an observer. The natural, coordinate induced orthonormal tetrad is
\begin{align}
\label{eq:tetrad}
\textbf{e}_{(0)}&=\partial_{t},&\quad \textbf{e}_{(1)}&=\partial_{\rho},\nonumber\\
\textbf{e}_{(2)}&=\frac{1}{r(\rho)}\partial_{\theta},&\quad \textbf{e}_{(3)}&=\frac{1}{r(\rho)\sin\theta}\partial_{\phi},
\end{align}
which satisfies
\begin{align}
g_{\mu\nu}e_{(a)}{}^{\mu}e_{(b)}{}^{\nu}=\eta_{(a)(b)}.
\end{align}
In the equatorial plane at some initial radius $\rho_{i}$, we construct such an orthonormal tetrad, then taking $\alpha\in(0,\pi)$ to be the angle the initial direction vector makes with the $\textbf{e}_{(1)}$-axis, an initial light-like direction (in units where $c=1$) takes the form
\begin{align}
\label{eq:lightdir}
\textbf{y}=\pm \textbf{e}_{(0)}+\cos\alpha \,\textbf{e}_{(1)}+\sin\alpha \,\textbf{e}_{(3)}
\end{align}
while an initial time-like direction may be written as
\begin{align}
\label{eq:timedir}
\textbf{y}=\pm\gamma\,\textbf{e}_{(0)}+v\,\gamma\,\cos\alpha\,\textbf{e}_{(1)}+v\,\gamma\,\sin\alpha\,\textbf{e}_{(3)},
\end{align}
where $v$ is the Euclidean norm of the three-velocity and $\gamma=1/\sqrt{1-v^{2}}$ is the Lorentz factor. The choice of sign in these expressions is determined by whether the geodesic is future or past-directed. It is straight-forward to check that
\begin{align}
\eta_{(a)(b)}y^{(a)}y^{(b)}=\varepsilon.
\end{align}
We can of course decompose the direction vector in the coordinate basis and use the geodesic equations to obtain
\begin{align}
\textbf{y}&=E\, \partial_{t}\pm\sqrt{E^{2}-V(L,\rho_{i})}\,\partial_{\rho}+\frac{L}{r^{2}(\rho_{i})}\,\partial_{\phi}\nonumber\\
&=E \,\textbf{e}_{(0)}\pm\sqrt{E^{2}-V(L,\rho_{i})} \,\textbf{e}_{(1)}+\frac{L}{r(\rho_{i})}\,\textbf{e}_{(3)}
\end{align}
where the latter equality follows from (\ref{eq:tetrad}). Comparison with Eqs.~(\ref{eq:lightdir})-(\ref{eq:timedir}) allows us to express the constants of motion in terms of the initial position and angle, \textit{viz}.,
\begin{align}
E&=\pm 1,\quad L=r(\rho_{i})\,\sin\alpha\quad\,\,\,\qquad\textrm{for null geodesics,}\nonumber\\
E&=\pm \gamma,\quad L=v\,\gamma \,r(\rho_{i})\,\sin\alpha\qquad\textrm{for time-like geodesics.}
\end{align}

\begin{figure*}
\centering
\includegraphics[width=18cm]{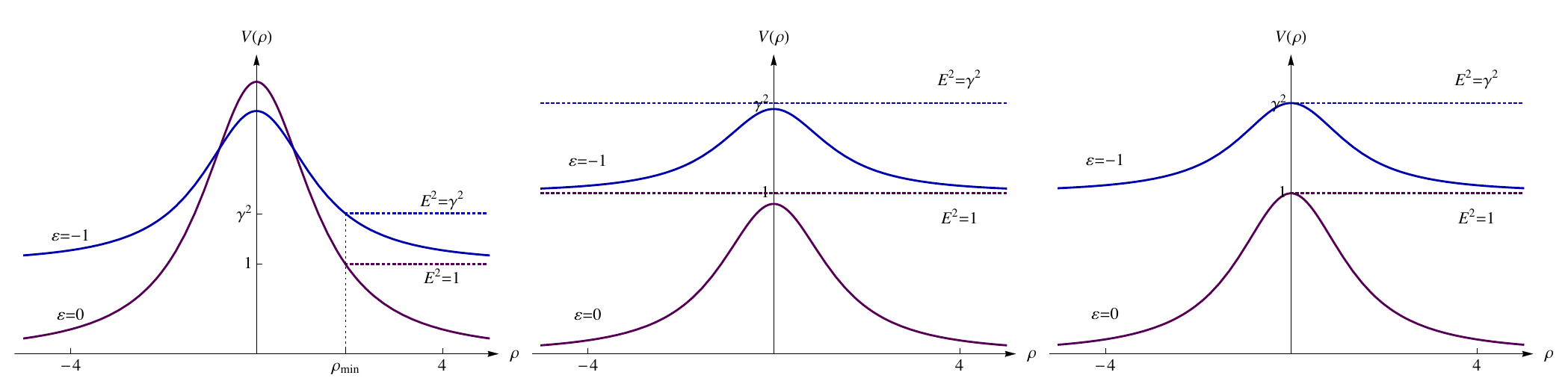}
\caption{{Plots of the scattering potentials for time-like (with $v=0.6$) and null geodesics. In these plots the geometry is held fixed with $b_{0}=1$ and $q=0$ and the initial position is fixed at $\rho_{i}=6$. We vary the initial angle $\alpha$. In the first plot $\alpha=0.35$ which implies $a<1$ and hence the geodesics are reflected by the potentials. In the second plot, we have $\alpha=2.95$ which corresponds to $a>1$ and hence the geodesic is transmitted through the wormhole throat to the lower universe. Finally, the third plot shows $\alpha=\alpha_{\textrm{crit}}$ for which the geodesic is just captured on an unstable bound orbit.} }
\label{fig:potentials}
\end{figure*}

Now the radial geodesic equation can be interpreted as a classical scattering problem with an angular momentum potential barrier $V(L,\rho)$ given by Eq.~(\ref{eq:potential}). Hence a geodesic can pass through the wormhole into the other universe if
\begin{align}
E^{2}>V(L,0)=\frac{L^{2}}{b_{0}^{2}}-\varepsilon,
\end{align}
or written in terms of the initial position and direction
\begin{align}
a^{2}\equiv\frac{b_{0}^{2}}{r^{2}(\rho_{i})\sin^{2}\alpha}>1.
\end{align}
We note that this condition is independent of whether the geodesic is null or time-like. Similarly, for a geodesic reflected back by the potential barrier, we have
\begin{align}
a^{2}<1.
\end{align}
In this case the proper radius of closest approach is given by the turning point $\dot{\rho}=0$, yielding the condition
\begin{align}
r(\rho_{\textrm{min}})=r(\rho_{i})\sin\alpha,
\end{align}
which can be substituted into Eq.~(\ref{eq:rho}) to give the explicit proper radius of closest approach,
\begin{align}
\label{eq:rhomin}
\rho_{\textrm{min}}=&\pm\frac{2 b_{0}i}{3-q}\Big(a^{(q-3)/2}{}_{2}F_{1}\Big(\frac{1}{2},\frac{3-q}{2(1-q)},\frac{5-3q}{2(1-q)},a^{q-1}\Big)\nonumber\\
&-\sqrt{\pi}\frac{\Gamma(\tfrac{5-3q}{2(1-q)})}{\Gamma(\tfrac{2-q}{1-q})}\Big).
\end{align}
The critical case $a=1$ has two solutions for the initial angle
\begin{align}
\alpha=\arcsin\frac{b_{0}}{r(\rho_{i})},\,\,\alpha=\pi-\arcsin\frac{b_{0}}{r(\rho_{i})},
\end{align}
representing an outgoing and ingoing geodesic, respectively. We denote the latter ingoing solution by $\alpha_{\textrm{crit}}$ and hence the outgoing solution is $\pi-\alpha_{\textrm{crit}}$ (Note that this is the opposite convention to Ref.~\cite{Muller}). The critical angle $\alpha_{\textrm{crit}}$ delimits between those geodesics that are reflected back by the potential and those which pass through to the other universe, the delimiting case corresponding to an unstable bound orbit. The three cases are illustrated in the effective potential plots in Fig.~\ref{fig:potentials}.

Dividing the radial and azimuthal geodesic equations permits us to obtain an equation for $r(\phi)$
\begin{align}
\label{eq:rphieqn}
\Big(\frac{dr}{d\phi}\Big)^{2}=\Big(1-\Big(\frac{b_{0}}{r}\Big)^{1-q}\Big)\Big[\frac{a^{2}}{b_{0}^{2}}r^{4}-r^{2}\Big].
\end{align}
Now let us define a dimensionless inverted radius by
\begin{align}
u=\frac{b_{0}}{a\,r(\rho)}=\frac{r(\rho_{i})}{r(\rho)}\sin\alpha,
\end{align}
which may be employed in Eq.~(\ref{eq:rphieqn}) yielding
\begin{align}
\Big(\frac{du}{d\phi}\Big)^{2}=(1-a^{1-q}u^{1-q})(1-u^{2}).
\end{align}
Hence we have a general solution of the form
\begin{align}
\label{eq:phiu}
\phi(u)=\pm\int_{\sin\alpha}^{u}\frac{dy}{\sqrt{(1-a^{1-q}y^{1-q})(1-y^{2})}},
\end{align}
or in terms of proper radial distance
\begin{align}
\label{eq:phirho}
\phi(\rho)=\pm\int_{\sin\alpha}^{(r(\rho_{i})/r(\rho))\sin\alpha}\frac{dy}{\sqrt{(1-a^{1-q}y^{1-q})(1-y^{2})}},
\end{align}
where the choice of sign is determined by whether the geodesic is ingoing ($d\rho<0$ which implies $du>0$) or outgoing ($d\rho>0$ which implies $du<0$). We have further used the fact that the initial position $u_{i}=u(\rho_{i})=\sin\alpha$. 

Unfortunately, this integral cannot be solved in terms of known functions for arbitrary $q$, and in general must be solved numerically. As before, we can numerically invert Eq.(\ref{eq:phirho}) to obtain $\rho(\phi)$ or $r(\phi)$ as needs be.

It is instructive to consider outgoing ($\alpha<\pi/2$) and ingoing ($\alpha>\pi/2)$ geodesics separately, which are further divided into those geodesics delimited by $a=1$.

\subsubsection{Case I: $0<\alpha<\pi-\alpha_{\textrm{crit}}$.}
For these outgoing geodesics, we have $a>1$. For numerical purposes, we would prefer if the integrand in Eq.~(\ref{eq:phiu}) was free from singularities. So let us define $\bar{a}=1/a<1$ and make the change of variable
\begin{align}
y=\bar{a}\sin w
\end{align}
which gives
\begin{align}
\phi(u)=\frac{-1}{a}\int_{\arcsin(a\,\sin\alpha)}^{\arcsin(a\,u)} \frac{\cos w\,dw}{\sqrt{(1-\sin^{1-q}w)(1-\bar{a}^{2}\sin^{2}w)}},
\end{align}
where the minus sign has been chosen to reflect the fact that the geodesic is outgoing. This angle reaches its maximum value as $\rho\to\infty$ ($u\to0$) and is given by
\begin{align}
\phi_{\textrm{max}}^{>}=\frac{1}{a}\int_{0}^{\arcsin(a\,\sin\alpha)} \frac{\cos w\,dw}{\sqrt{(1-\sin^{1-q}w)(1-\bar{a}^{2}\sin^{2}w)}}.
\end{align}

\subsubsection{Case II: $\pi-\alpha_{\textrm{crit}}\le\alpha\le\pi/2$.}
For initial angles in this range, we have $a\le1$. Here we make a change of variable $y=\sin w$, we arrive at the particularly simple expression
\begin{align}
\phi(u)=-\int_{\alpha}^{\arcsin(u)}\frac{dw}{\sqrt{1-a^{1-q}\sin^{1-q}w}},
\end{align}
which asymptotes to the maximum value as $\rho\to\infty$,
\begin{align}
\phi_{\textrm{max}}^{<}=\int_{0}^{\alpha}\frac{dw}{\sqrt{1-a^{1-q}\sin^{1-q}w}}.
\end{align}
For the critical case $\alpha=\pi-\alpha_{\textrm{crit}}=\arcsin(b_{0}/r(\rho_{i}))$, we have $a=1$ and $u=b_{0}/r(\rho)$. Hence the angle is given by 
\begin{align}
\phi_{\textrm{crit}}^{+}(u)=-\int_{\arcsin(b_{0}/r(\rho_{i}))}^{\arcsin(b_{0}/r(\rho))}\frac{dw}{\sqrt{1-\sin^{1-q}w}}.
\end{align}

\subsubsection{Case III: $\pi/2<\alpha\le\alpha_{\textrm{crit}}$.}
For these initially ingoing geodesics, we have $a\le1$ and the potential barrier implies that we must have $\rho\ge\rho_{\textrm{min}}$ where $\rho_{\textrm{min}}$ is given by Eq.~(\ref{eq:rhomin}). For an ingoing geodesic, since the initial angle is in the second quadrant of the unit circle, we adopt the transformation
\begin{align}
w=\pi-\arcsin(y),
\end{align}
which is a monotonically decreasing function of $w$ and hence
\begin{align}
dw=-\frac{dy}{\sqrt{1-y^{2}}}=\frac{dy}{\cos w},
\end{align}
where the last equality holds since $\cos w<0$ in the second quadrant. Hence for the ingoing branch of the integral representation (\ref{eq:phiu}), we arrive at
\begin{align}
\label{eq:philess1in}
\phi(u)=-\int_{\alpha}^{\pi-\arcsin(u)}\frac{dw}{\sqrt{1-a^{1-q}\sin^{1-q}w}}.
\end{align}
We also have geodesics which are initially ingoing, reach the minimum value permitted by the potential barrier $u(\rho_{\textrm{min}})=1$ and are then reflected and begin to recede to spatial infinity. For the reflected outgoing portion of the geodesic trajectory we employ the substitution $w=\arcsin(y)$ in the negative branch of Eq.~(\ref{eq:phiu}). Putting these together yields
\begin{align}
\label{eq:philess1out}
\phi(u)=&-\int_{\alpha}^{\pi/2}\frac{dw}{\sqrt{1-a^{1-q}\sin^{1-q}w}}\nonumber\\
&-\int_{\pi/2}^{\arcsin(u)}\frac{dw}{\sqrt{1-a^{1-q}\sin^{1-q}w}},
\end{align}
where the first term is the contribution coming from the ingoing geodesic integrating down to the minimum proper radius allowed by the potential barrier, while the the second integral is the contribution from the outgoing geodesic. This is most succinctly expressed as the single integral
\begin{align}
\phi(u)=-\int_{\alpha}^{\arcsin(u)}\frac{dw}{\sqrt{1-a^{1-q}\sin^{1-q}w}}.
\end{align}
For these reflected geodesics, since $u(\rho_{\textrm{min}})=1$ at the point of closest approach, the angle at this turning point is given by 
\begin{align}
\phi_{\textrm{turning}}=-\int_{\alpha}^{\pi/2}\frac{dw}{\sqrt{1-a^{1-q}\sin^{1-q}w}},
\end{align}
while the maximum value as the geodesic recedes to infinity is given by
\begin{align}
\phi_{\textrm{max}}^{\textrm{ref}}=\int_{0}^{\alpha}\frac{dw}{\sqrt{1-a^{1-q}\sin^{1-q}w}}.
\end{align}

For the critical ingoing geodesics, we have $\alpha=\alpha_{\textrm{crit}}$ and $a=1$, which describes those unstable bound geodesics captured on the throat of the wormhole. Such geodesics are given by
\begin{align}
\phi_{\textrm{crit}}^{-}(u)=-\int_{\pi-\arcsin(b_{0}/r(\rho_{i}))}^{\pi-\arcsin(b_{0}/r(\rho))}\frac{dw}{\sqrt{1-\sin^{1-q}w}}.
\end{align}

\begin{figure*}
\centering
\includegraphics[width=18cm]{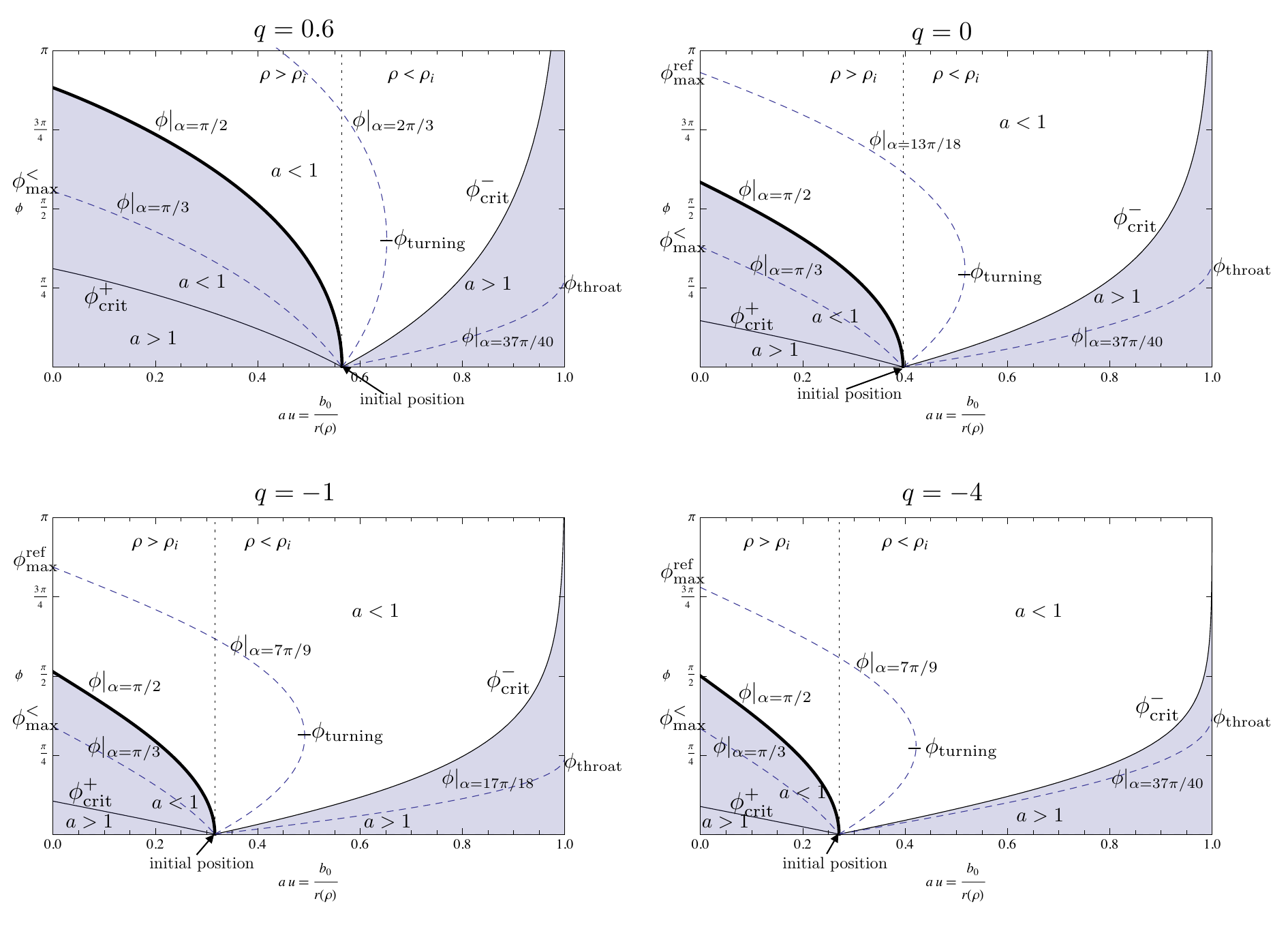}
\caption{{Plots of the the geodesics in $(a\,u,\phi)$ coordinate space for various values of the shape exponent $q$. In these plots the throat width has been set to $b_{0}=2$ and the initial position is fixed at $\rho_{i}=6$. In each graph, several geodesics are plotted with different initial angle $\alpha$, which determines whether the geodesic recedes to spatial infinity, traverses the wormhole or is initially ingoing and then is reflected to spatial infinity. The thick curve is the geodesic with initial angle $\alpha=\pi/2$ which separates geodesics which are initially ingoing from those that are initially outgoing. The left shaded region in each plot is the space of initially outgoing geodesics which simply recede to spatial infinity. The initially ingoing geodesics are further divided into those that are reflected back to spatial infinity (the unshaded area) and those which traverse the wormhole (the right shaded region). The only geodesic which does not end up at spatial infinity ($\rho\pm\infty$) is the unstable bound orbit at $\rho=0$ which is shown as $\phi_{\textrm{crit}}^{-}$ in the plots above. In terms of the variation with the shape exponent $q$, it is clear from the plots that for fixed throat radius and initial position, the space of initially ingoing geodesics and the space of traversing geodesics decreases as $q$ decreases, while the space of reflected geodesics is increasing as $q$ decreases. In other words, if one emitted a light ray from a certain position $\rho_{i}$ in a random direction in a wormhole space-time with throat radius $b_{0}$ and shape exponent $q$, then the chances of that light ray traversing the wormhole would be less than if the light ray had been emitted in a wormhole with a more positive shape exponent, assuming $b_{0}$ and $\rho_{i}$ the same.} }
\label{fig:PhiPlots}
\end{figure*} 

\begin{figure*}
\centering
\includegraphics[width=18cm]{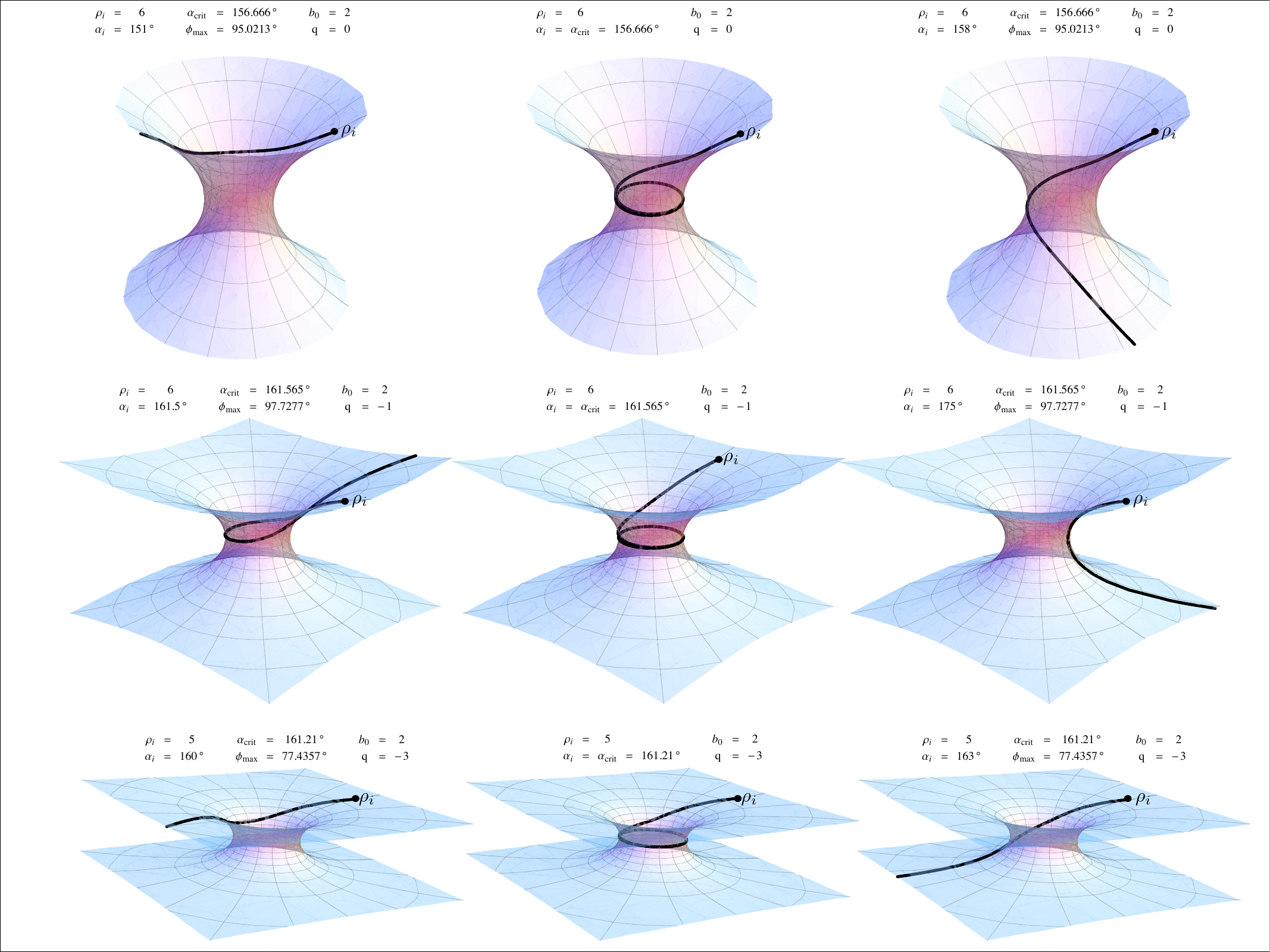}
\caption{{Embedding plots of the geodesic trajectories for various shape exponents and various initial conditions. For each wormhole geometry, we give an example of a reflected, a trapped and a transmitted geodesic.} }
\label{fig:geodesicplots}
\end{figure*} 

\subsubsection{Case IV: $\alpha>\alpha_{\textrm{crit}}$.}
For the initial direction in this region, we have $a>1$. For the ingoing geodesic in the upper universe ($\rho>0$), analogous to the previous case, we employ the substitution
\begin{align}
 w=\pi-\arcsin(a\,y)\quad\implies\quad dw=\frac{a}{\cos w}dy,
 \end{align}
 in Eq.~(\ref{eq:phiu}), where again we note that $\cos w<0$ in the second quadrant. Hence for geodesics that remain in the same universe, the integral becomes
\begin{align}
\phi(u)=\frac{1}{a}\int_{\pi-\arcsin(a\,\sin\alpha)}^{\pi-\arcsin(a\,u)} \frac{\cos w\,dw}{\sqrt{(1-\sin^{1-q}w)(1-\bar{a}^{2}\sin^{2}w)}}.
\end{align}
For geodesics that traverse the wormhole, we make the change of variables $w=\arcsin(y)$ for the outgoing transmitted geodesic in the lower universe. We obtain
\begin{align}
\phi(u)=&\frac{1}{a}\int_{\pi-\arcsin(a\,\sin\alpha)}^{\pi/2}  \frac{\cos w\,dw}{\sqrt{(1-\sin^{1-q}w)(1-\bar{a}^{2}\sin^{2}w)}}\nonumber\\
&-\frac{1}{a}\int_{\pi/2}^{\arcsin(a\,u)}  \frac{\cos w\,dw}{\sqrt{(1-\sin^{1-q}w)(1-\bar{a}^{2}\sin^{2}w)}},
\end{align}
or equivalently
\begin{align}
\phi(u)=&\frac{1}{a}\int_{\arcsin(a\,\sin\alpha)}^{\pi/2}  \frac{\cos w\,dw}{\sqrt{(1-\sin^{1-q}w)(1-\bar{a}^{2}\sin^{2}w)}}\nonumber\\
&-\frac{1}{a}\int_{\pi/2}^{\arcsin(a\,u)}  \frac{\cos w\,dw}{\sqrt{(1-\sin^{1-q}w)(1-\bar{a}^{2}\sin^{2}w)}}.
\end{align}
The throat crossing occurs at $u(\rho=0)=1/a$, whence the angle is given by
\begin{align}
\phi_{\textrm{throat}}=\frac{1}{a}\int_{\arcsin(a\,\sin\alpha)}^{\pi/2}  \frac{\cos w\,dw}{\sqrt{(1-\sin^{1-q}w)(1-\bar{a}^{2}\sin^{2}w)}}.
\end{align}
Moreover, these geodesics that traverse the wormhole asymptote to their maximum value as $\rho\to-\infty$,
\begin{align}
\phi_{\textrm{max}}^{\textrm{trav}}=&\frac{1}{a}\int_{\arcsin(a\,\sin\alpha)}^{\pi/2}  \frac{\cos w\,dw}{\sqrt{(1-\sin^{1-q}w)(1-\bar{a}^{2}\sin^{2}w)}}\nonumber\\
&+\frac{1}{a}\int_{0}^{\pi/2} \frac{\cos w\,dw}{\sqrt{(1-\sin^{1-q}w)(1-\bar{a}^{2}\sin^{2}w)}}.
\end{align}
These solutions to the geodesic equations have been plotted in $(a\,u,\phi)$ coordinates for various shape exponents in Fig.~\ref{fig:PhiPlots}.

Now the most natural way to visualize the actual geodesic trajectories in the wormhole space-time is by an embedding in Euclidean space in cylindrical coordinates. For the wormhole space-time itself, we simply rotate the embedding function $z(r)$ about the vertical axis to give us a natural representation of a static equatorial slice of the wormhole geometry. In order to plot the geodesics propagating on this background, we require a parametric representation of the solutions of the geodesic equations in cylindrical coordinates. To achieve this, we can numerically invert the solutions above for $\phi(u)$ in order to obtain $u(\phi)$ and hence $r(\phi)=b_{0}/(a u(\phi))$. For example, if we take an initially ingoing geodesic with initial angle $\pi/2<\alpha<\alpha_{\textrm{crit}}$, then the geodesic is reflected and recedes to infinity. We can then invert Eq.~(\ref{eq:philess1in}) and (\ref{eq:philess1out}) by
\begin{align}
\label{eq:uless1ref}
u(\phi)=\begin{cases}
\Big\{X: -\displaystyle{\int_{\alpha}^{\pi-\arcsin(X)}\frac{dw}{\sqrt{1-a^{1-q}\sin^{1-q}w}}}=\phi\Big\}, \\ \\
\qquad\qquad\qquad\qquad\qquad\qquad\quad0<\phi<\phi_{\textrm{turning}}\\ \\
\Big\{X: -\displaystyle{\int_{\alpha}^{\pi/2}\frac{dw}{\sqrt{1-a^{1-q}\sin^{1-q}w}}}\\
\qquad\,\,-\displaystyle{\int_{\pi/2}^{\arcsin(X)}\frac{dw}{\sqrt{1-a^{1-q}\sin^{1-q}w}}}=\phi\Big\},\\ \\
 \qquad\qquad\qquad\qquad\qquad\,\,\,\,\,\,\,\,\phi_{\textrm{turning}}<\phi<\phi_{\textrm{max}}^{\textrm{ref}},
\end{cases}
\end{align}
which gives the ingoing and reflected branch of the curve, respectively. Finally to plot this geodesic in the embedding space, each point along the curve is assigned the parametric representation $(r(\phi)\cos\phi,r(\phi)\sin\phi, z(r(\phi)))$ with $r(\phi)=b_{0}/(a u(\phi))$, where in the particular case of the reflected geodesic, $u(\phi)$ is given by Eq.~(\ref{eq:uless1ref}) and the parameter $\phi$ runs over the interval $[0,\phi_{\textrm{max}}^{\textrm{ref}})$. Analogous statements can be made for the other types of geodesics. The result of these parametric plots are shown in Fig.~\ref{fig:geodesicplots} for a range of different shape exponents and different initial conditions.

\section{Scalar Field Propagation}
\label{sec:scalar}
We now consider a massless scalar field $\varphi(x)$ propagating on the background space-time (\ref{eq:metric}) where the scalar is non-minimally coupled to the gravitational field with coupling strength $\xi$. Hence $\varphi(x)$ satisfies the wave equation
\begin{align}
\label{eq:waveeqn}
(\Box-\xi\, R)\varphi(x)=0,
\end{align}
where $R$ is the Ricci curvature scalar, which for arbitrary shape function $b(r)$ is given by
\begin{align}
R=\frac{2b'(r)}{r^{2}}.
\end{align}
Written explicitly in $(t,r,\theta,\phi)$ coordinates, the wave equation is
\begin{align}
&\Big\{-\frac{\partial^{2}}{\partial t^{2}}+\frac{1}{r^{2}}\Big(1-\frac{b(r)}{r}\Big)^{1/2}\frac{\partial}{\partial r}\Big(r^{2}\Big(1-\frac{b(r)}{r}\Big)^{1/2}\frac{\partial}{\partial r}\Big)\nonumber\\
&+\frac{1}{r^{2}\sin\theta}\frac{\partial}{\partial\theta}\Big(\sin\theta \frac{\partial}{\partial\theta}\Big)+\frac{1}{r^{2}\sin^{2}\theta}\frac{\partial^{2}}{\partial\phi^{2}}\nonumber\\
&-2\xi\Big(\frac{b'(r)}{r^{2}}\Big)\Big\}\varphi(x)=0.
\end{align}
Solutions can be obtained by a separation of variables and are given by
\begin{align}
\varphi(x)=e^{-i\omega t}e^{i m \phi}P_{l}^{m}(\cos\theta)\chi_{\omega l}(r)/r,
\end{align}
where $P_{l}^{m}(\cos\theta)$ are the associated Legendre functions, and $\chi_{\omega l}(r)$ satisfies the ordinary differential equation
\begin{align}
\Big\{\Big(1-\frac{b(r)}{r}\Big)^{1/2}\frac{d}{dr}\Big(\Big(1-\frac{b(r)}{r}\Big)^{1/2}\frac{d}{dr}\Big)+\omega^{2}-\frac{l(l+1)}{r^{2}}\nonumber\\
-2(\xi-\tfrac{1}{4})\frac{b'(r)}{r^{2}}-\frac{b(r)}{2 r^{3}}\Big\}\chi_{\omega l}(r)=0.
\end{align}
Rewriting in terms of proper radial distance $\rho$ yields
\begin{align}
\label{eq:wavepropeqn}
\Big\{\frac{d^{2}}{d\rho^{2}}+\omega^{2}-V_{l}(\xi,\rho)\Big\}\chi_{\omega l}(\rho)=0,
\end{align}
where the potential is given by
\begin{align}
	\label{eq:radialpotential}
V_{l}(\xi,\rho)=\frac{l(l+1)}{r^{2}}+2(\xi-\tfrac{1}{4})\frac{b'(r)}{r^{2}}+\frac{b(r)}{2 r^{3}}.
\end{align}
This potential can be very different from those that arise in black hole space-times which are everywhere positive with a single peak. That the second term in the potential can be negative for wormhole spacetimes can give rise to a rich potential structure depending on the mode, the specific shape function and the coupling strength. For example, we can have potentials with multiple turning points or negative definite potentials. The latter case is typically indicative of an instability which we investigate further in the following subsection.

\subsection{Stability}
\label{sec:stability}
If we turn our attention back to the particular class of wormholes defined by the shape functions given in Eq.~(\ref{eq:shapefn}), then the Ricci scalar is given by
\begin{align}
R=\frac{2 q}{r^{2}}\Big(\frac{b_{0}}{r}\Big)^{1-q},
\end{align}
and the potential (\ref{eq:radialpotential}) simplifies to
\begin{align}
	\label{eq:potentialbr}
	V_{l}(\xi,\rho)=\frac{1}{r^{2}}\Big(l(l+1)+\Big(\frac{b_{0}}{r}\Big)^{1-q}\{2 q \,\xi+\tfrac{1}{2}(1-q)\}\Big).
\end{align}
When this potential is positive definite, the operator
\begin{align}
A=-\frac{d^{2}}{d\rho^{2}}+V
\end{align}
is a positive self-adjoint operator on the Hilbert space of square integrable functions of $\rho$. Wald \cite{Wald1979} has proven that for such operators, given some well-behaved initial data, solutions to the wave equation remain bounded for all time. Hence, it is clear from (\ref{eq:potentialbr}) that the scalar field $\varphi$ is stable whenever $2 q \xi+\tfrac{1}{2}(1-q)>0$. Solving this inequality for $\xi$ implies that
\begin{align}
	\label{eq:stablexi}
	\xi &>\frac{q-1}{4q},\qquad 0<q<1\nonumber\\
	\xi &<\frac{q-1}{4q},\qquad q<0.
\end{align}
These inequalities imply, for example, that all scalar solutions, regardless of shape exponent $q<1$, are stable if the coupling constant is in the range $0\le\xi\le\tfrac{1}{4}$. We also note that for $q=0$, the Ricci scalar vanishes and hence $\xi$ does not appear in the potential. In this case, the scalar field solutions are stable since the potential is clearly positive definite. 

However, it is also clear that there is a region of the parameter space for which the potential (\ref{eq:potentialbr}) is not positive definite and Wald's argument breaks down. For example, for $q<0$ and $\xi>(q-1)/4q$, the $l=0$ mode gives rise to a negative definite potential and hence this mode is always unstable. Indeed the $l=0$ mode places the strongest constraint on $\xi$ for stable solutions. In Sec.~\ref{sec:qnm}, we compute the WKB approximation for the quasi-normal modes and show explicitly that there are certain values of the coupling constant that yield unstable modes.

\begin{figure}
\centering
\includegraphics[width=6.7cm]{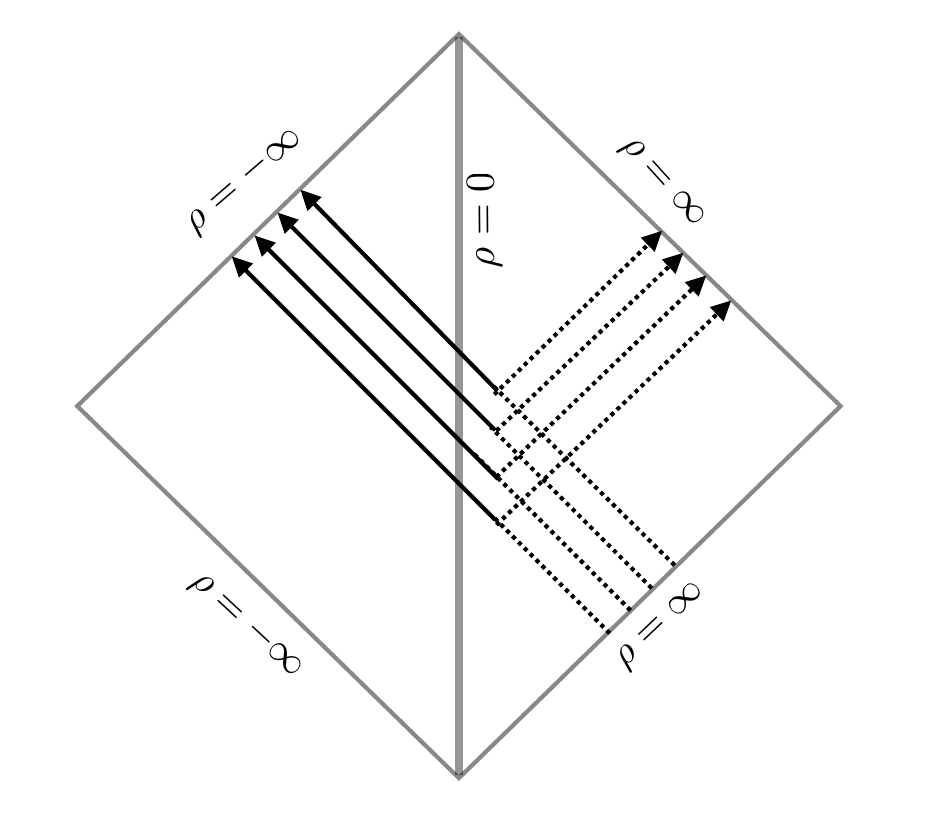}
\caption{{Penrose diagram for a wormhole connecting two asymptotically flat universes. The diagram shows incoming radiation from spatial infinity $\rho=\infty$ being partially reflected back to infinity and partly transmitted through the wormhole throat to reach the other spatial infinity $\rho=-\infty$.} }
\label{fig:penrose}
\end{figure}

\begin{figure*}
\centering
\includegraphics[width=18.5cm]{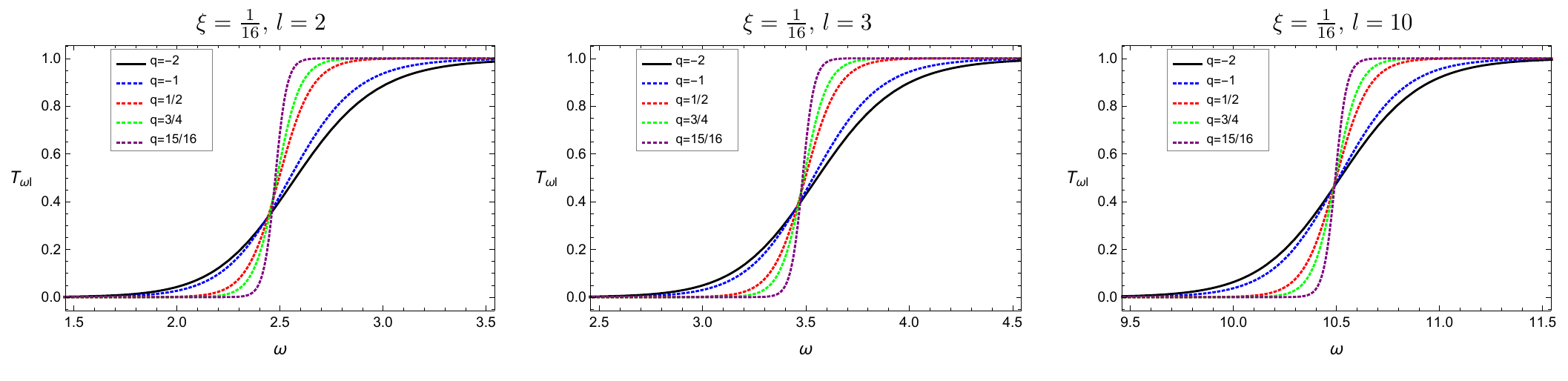}
\caption{{Plot of transmission coefficients for various shape exponents for $l=2,3,10$. The coupling constant has been set to the conformal value $\xi=1/16$ and the throat radius is set to unity.} }
\label{fig:tranvarq}
\end{figure*} 

\begin{figure*}
\centering
\includegraphics[width=18.5cm]{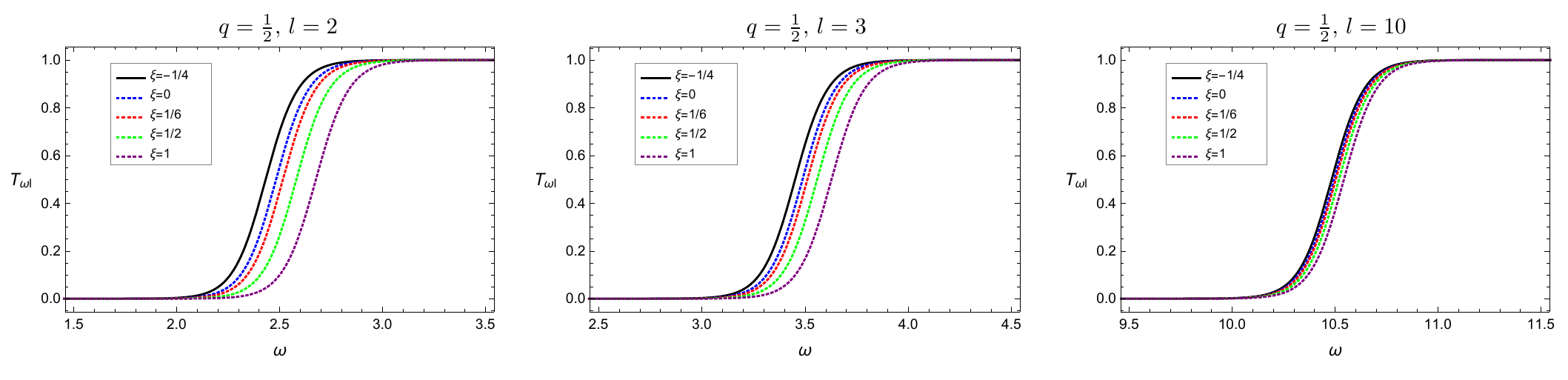}
\caption{{Plot of transmission coefficients for various values of the coupling constant for $l=2,3,10$. The shape exponent is $q=1/2$ and the throat radius is set to unity.} }
\label{fig:tranvarxi}
\end{figure*} 

\subsection{Computation of Reflection and Transmission Coefficients}
For stable values of $\xi$ given by (\ref{eq:stablexi}), the potential is everywhere positive definite and asymptotes to zero at both spatial infinities. Hence we have a typical wave scattering problem analogous to black hole scattering. We imagine an ingoing (moving towards the wormhole throat) wave from spatial infinity scattering off the potential barrier, some portion of the radiation is reflected back to spatial infinity and some portion is transmitted through the wormhole throat and reaches the other spatial infinity. This is represented in the Penrose diagram in Fig.~\ref{fig:penrose}. Hence the appropriate boundary conditions are
\begin{align}
\chi_{\omega l}&\sim e^{-i\omega \rho}+A^{\textrm{ref}}_{\omega l}e^{i\omega\rho}&\qquad\rho&\to\infty,\nonumber\\
\chi_{\omega l}&\sim A^{\textrm{tran}}_{\omega l}e^{-i\omega\rho} &\qquad\rho&\to-\infty,
\end{align}
where $A_{\omega l}^{\textrm{ref/tran}}$ are the amplitudes of the reflected/transmitted waves, respectively, which are related to the reflection/transmission coefficients by
\begin{align}
T_{\omega l}\equiv & |A^{\textrm{tran}}_{\omega l}|^{2}\nonumber\\
R_{\omega l}\equiv & |A^{\textrm{ref}}_{\omega l}|^{2}=1-T_{\omega l}.
\end{align}
To compute the reflection and transmission coefficients to high accuracy, we adopt the fourth order WKB approximation derived by Will and Guinn \cite{WillGuinn} which is based on the third order WKB approximation developed by Iyer and Will \cite{IyerWill1} in order to compute black hole normal modes. The method involves simultaneously matching the asymptotic WKB series to a Taylor expansion about the peak of the potential and hence is valid for frequencies $\omega^{2}\approx V_{0}$. This has been extended to higher orders and applied to the cases of minimally coupled scalar and electromagnetic fields \cite{KonoplyaZhidenko} in the same class of wormholes under consideration here. The WKB method results in a transmission coefficient
\begin{align}
T_{\omega l}=(1+e^{2\pi i S})^{-1},
\end{align}
where $S$ is a pure imaginary factor defined implicitly by
\begin{align}
S=i \frac{(\omega^{2}-V_{0})}{\sqrt{-2V_{0}''}}-\Lambda^{(2)}-\Lambda^{(3)}-\Lambda^{(4)}.
\end{align}
The $\Lambda^{(i)}$ are higher order WKB corrections which themselves contain a dependence on $S$ up to quadratic order. This is not an exact equation to be solved as a polynomial in $S$, but rather the $\Lambda^{(i)}$ are progressively smaller corrections and hence we solve for $S$ by iteration. For notational convenience, we introduce the variables
\begin{align}
k=-2 V_{0}^{''},\qquad q_{0}^{2}=\frac{\omega^{2}-V_{0}}{V_{0}^{''}}>0,\qquad v_{(n)}=\frac{V_{0}^{(n)}}{V_{0}^{''}},
\end{align}
where subscript zero denotes evaluation at the wormhole throat $\rho=0$ which coincides with the maximum of the potential. As in the geodesic equation, the potential is symmetric about the throat and hence $v_{(n)}\equiv0$ for all odd orders. The WKB corrections are explicitly given by
\begin{align}
\Lambda^{(2)}&=\frac{i}{8\sqrt{k}}v_{(4)}(\tfrac{1}{4}+S^{2})\nonumber\\
\Lambda^{(3)}&=\frac{S}{k}\left\{\tfrac{67}{2304}v_{(4)}^{2}-\tfrac{5}{288}v_{(6)}\right\}+\frac{S^{3}}{k}\left\{\tfrac{17}{576}v_{(4)}^{2}-\tfrac{1}{72}v_{(6)}\right\}\nonumber\\
\Lambda^{(4)}&=-\frac{1}{k^{3/2}}\left\{\tfrac{57}{16384}v_{(4)}^{3}-\tfrac{7}{2048}v_{(4)}v_{(6)}+\tfrac{1}{2048}v_{(8)}\right\}\nonumber\\
&-\frac{S^{2}}{k^{3/2}}\left\{\tfrac{569}{18432}v_{(4)}^{3}-\tfrac{59}{2304}v_{(4)}v_{(6)}+\tfrac{7}{2304}v_{(8)}\right\}\nonumber\\
&-\frac{S^{4}}{k^{3/2}}\left\{\tfrac{125}{9216}v_{(4)}^{3}-\tfrac{11}{1152}v_{(4)}v_{(6)}+\tfrac{1}{1152}v_{(8)}\right\}.
\end{align}
Repeated substitution of $S$ into these expressions and only retaining terms to the required order gives the following solution for $S$:
\begin{align}
\label{eq:Swkb}
S=-i k^{1/2}\Big\{\tfrac{1}{2}q_{0}^{2}-\tfrac{1}{32}v_{(4)}q_{0}^{4}+\left(\tfrac{35}{4608}v_{(4)}^{2}-\tfrac{1}{576}v_{(6)}\right)q_{0}^{6}\nonumber\\
 +\left(-\tfrac{385}{147456}v_{(4)}^{3}+\tfrac{7}{6144}v_{(4)}v_{(6)}-\tfrac{1}{18432}v_{(8)}\right)q_{0}^{8}\Big\}\nonumber\\
 -i k^{-1/2}\Big\{\tfrac{1}{32}v_{(4)}-\left(\tfrac{85}{4608}v_{(4)}^{2}-\tfrac{5}{576}v_{(6)}\right)q_{0}^{2}\nonumber\\
 -\left(-\tfrac{875}{73728}v_{(4)}^{3}+\tfrac{77}{9216}v_{(4)}v_{(6)}-\tfrac{7}{9216}v_{(8)}\right)q_{0}^{4}\Big\}\nonumber\\
 -ik^{-3/2}\Big\{-\tfrac{665}{147456}v_{(4)}^{3}+\tfrac{73}{18432}v_{(4)}v_{(6)}-\tfrac{1}{2048}v_{(8)}\Big\}.
\end{align}
We have plotted various transmission coefficients for different values of $q$ and $\xi$ in Figs.~\ref{fig:tranvarq}-\ref{fig:tranvarxi}. We note a few key features of these plots. First, as noted in Ref.~\cite{KonoplyaZhidenko}, we see from Fig.~\ref{fig:tranvarq} that the transmission coefficients approach the step function $\theta(\omega^{2} b_{0}^{2}-l(l+1))$ as $q\to 1$. Even away from this limit, it is clear that the transmission coefficient is non-trivial only in a small region about $\omega\approx l$. For fixed $\xi$, the effect of increasing $l$ results in a simple linear shift in the transmission coefficient that is only weakly dependent on the shape function $q$. We can see from the graphs also that for fixed $\xi$ and $l$, each graph for various $q$ approximately passes through a common point, and this point corresponds to the reflection coefficient at the peak of the potential $\omega^{2}=V$. In the large $l$ limit, the common point is exactly the frequency at the peak of the potential and the reflection and transmission coefficients are equal, i.e., half the amplitude is transmitted and half reflected. To see this explicitly, we note that at the peak of the potential $q_{0}^{2}=0$ and hence only the term proportional to $k^{-3/2}$ survives in Eq.~(\ref{eq:Swkb}). It is straight-forward to show that this term vanishes in the large $l$ limit and hence $T_{\omega l}=R_{\omega l}=1/2$. 

Turning now to Fig.~\ref{fig:tranvarxi}, we see that for fixed $l$ and $q$, a change in the coupling strength $\xi$ induces a shift in the graph of the transmission coefficient, and hence the graphs do not cross, i.e., the transmission coefficient is a slowly monotonically increasing function of the scalar coupling. On the other hand, we see from the coalescing of the graphs that the dependence on the coupling is suppressed by larger $l$ modes. This is not surprising of course, since in the large $l$ geometric optics limit, the terms that are independent of the principal part of the wave equation are subdominant, or stated another way, the behavior of the field on very small length-scales is dominated by the terms in the wave equation that involve second-order derivatives.

 \begin{table*}[]
 \begin{center}
\begin{tabular}{| c | c || c | c | c | c |} \hline
 &  &$\xi=0$ & $\xi=\tfrac{1}{8}$ &  $\xi=\tfrac{1}{6}$  & $\xi=\tfrac{31}{64}$   \\[5pt]
\hline\hline
 \multirow{2}{*}{$l=1$}  & $n=0$ & $1.54470 - 0.56885 i$ & $1.46687 - 0.55894 i $ &  $1.44057 - 0.55550 i$ &  $1.241469 - 0.520920 i$ \\ 
   &  $n=1$  & $ 1.25818 - 1.80846 i$ & $ 1.17398 - 1.78376 i $ &  $1.14745 - 1.77481 i$ &  $1.00524 - 1.66915 i$ \\[5pt] \hline
  \multirow{3}{*}{$l=2$} & $n=0$  & $ 2.53614 - 0.52511 i $ & $2.48862 - 0.51792 i $ &  $2.47267 - 0.51539 i$ &  $2.34961 - 0.49322 i$  \\
   &  $n=1$ &$2.32911 - 1.62589 i$ &  $ 2.28098 - 1.60558 i$ & $2.26509 - 1.59840 i$ &  $2.14771 - 1.53321 i$  \\
   &  $n=2$  & $ 1.94921 - 2.78714 i $ &  $1.90192 - 2.75542 i $ & $1.88693 - 2.74412 i$ &  $1.78585 - 2.63785 i$ \\[5pt] \hline 
   \multirow{4}{*}{$l=3$} & $n=0$  & $ 3.52966 - 0.51194 i $ & $3.49514 - 0.50760 i $ & $3.48358 - 0.50611 i$ &  $3.39459 - 0.49393 i$  \\
   &  $n=1$ &$3.37893 - 1.56045 i$ &  $3.34463 - 1.54754 i $ &$3.33321 - 1.54309 i$ &  $3.24666 - 1.50640 i$  \\
   &  $n=2$  & $ 3.09518 - 2.65077 i$ & $3.06160 - 2.62965 i $ & $3.05056 - 2.62235 i$ &  $2.96916 - 2.56123 i$ \\
   & $n=3$ & $ 2.67257 - 3.77252 i   $   &  $2.64096 - 3.74384 i $ & $ 2.63078 - 3.73387 i  $   &   $ 2.55867 - 3.64885 i  $  \\[5pt] \hline
    \multirow{5}{*}{$l=4$} & $n=0$  & $ 4.52468 - 0.50680 i$ & $ 4.49754 - 0.50400 i$ &  $4.48846 - 0.50305 i$ &  $4.41878 - 0.49550 i$  \\
   &  $n=1$ &$4.40830 - 1.53414 i$ & $4.38144 - 1.52573 i $ & $4.37248 - 1.52287 i$ &  $4.30414 - 1.50008 i$  \\
   &  $n=2$  & $ 4.18431 - 2.58875 i$ & $4.15802 - 2.57474 i $ & $4.14930 - 2.56997 i$ &  $4.08360 - 2.53175 i$ \\
   & $n=3$ & $ 3.85419 - 3.66838 i  $   & $3.82892 - 3.64893 i $ &  $  3.82060 - 3.64229 i $   &   $ 3.75906 - 3.58866 i  $ \\
   & $n=4$ & $ 3.40931 - 4.77191 i   $   &  $3.38586 - 4.74728 i $ &  $ 3.37822 - 4.73884 i  $   &   $ 3.32289 - 4.66993 i  $  \\[5pt] \hline
    \end{tabular}
\end{center}
\caption{We tabulate the quasi-normal modes for mode numbers $l=1$ to $l=4$ for various values of the coupling constant in the Ellis ($q=-1$) wormhole. The throat radius has been set to unity.}
\label{tab:qnmtable}
\end{table*}

\subsection{Quasi-Normal Modes}
\label{sec:qnm}
The quasi-normal modes (QNMs) of a scalar field on a wormhole space-time are the complex frequencies corresponding to pure outgoing radiation at both spatial infinities, where the real part of the mode determines the oscillation frequency and the imaginary part determines the damping. The WKB method was adopted in Ref.~\cite{Kim} to compute QNMs for minimal coupling for the $q=-1$ case and extended to arbitrary $q$ and higher-orders in the WKB expansion in Ref.~\cite{KonoplyaZhidenko}. Here we extend these results to include non-minimal coupling. Again we adopt the fourth-order WKB method \cite{IyerWill1} which gives the square of the quasi-normal mode frequency
\begin{align}
\omega^{2}&=\left(V_{0}+\tfrac{1}{8}v_{(4)}(\tfrac{1}{4}+(n+\tfrac{1}{2})^{2})\right)\nonumber\\
&+i\Big[\frac{1}{k}\left(\tfrac{57}{16384}v_{(4)}^{3}-\tfrac{7}{2048}v_{(4)}v_{(6)}+\tfrac{1}{2048}v_{(8)}\right)\nonumber\\
&-\frac{(n+\tfrac{1}{2})}{k^{1/2}}\left(k+\tfrac{67}{2304}v_{(4)}^{2}-\tfrac{5}{288}v_{(6)}\right)\nonumber\\
&+\frac{(n+\tfrac{1}{2})^{2}}{k}\left(\tfrac{569}{18432}v_{(4)}^{3}-\tfrac{59}{2304}v_{(4)}v_{(6)}+\tfrac{7}{2304}v_{(8)}\right)\nonumber\\
&-\frac{(n+\tfrac{1}{2})^{3}}{k^{1/2}}\left(\tfrac{17}{576}v_{(4)}^{2}-\tfrac{1}{72}v_{(6)}\right)\nonumber\\
&+\frac{(n+\tfrac{1}{2})^{4}}{k}\left(\tfrac{125}{9216}v_{(4)}^{3}-\tfrac{11}{1152}v_{(4)}v_{(6)}+\tfrac{1}{1152}v_{(8)}\right)\Big].
\end{align}
We have tabulated some values for the first few $l$-modes with $q=-1$ for various values of the coupling constant in Table~\ref{tab:qnmtable}. The first thing to note from these values is that the imaginary part is always negative which rules out any exponentially growing modes as expected since we have limited ourselves to the stable domain $\xi<1/2$. In the large $l$ limit, we find
\begin{align}
\textrm{Re}(\omega)&=\frac{1}{b_{0}}(l+\tfrac{1}{2})-\frac{1}{128 b_{0}(l+\tfrac{1}{2})}\Big[16-128\xi\nonumber\\
&+(1-q)(128\xi-22)+(1-q)^{2}\nonumber\\
&+4 (n+\tfrac{1}{2})^{2}(1-q)(3-q)\Big]+\mathcal{O}(l^{-3}),\nonumber\\
\textrm{Im}(\omega)&=-\frac{(n+\tfrac{1}{2})\sqrt{1-q}}{\sqrt{2} b_{0}}\Big\{1-\frac{(1-q)}{6144 b_{0}(l+\tfrac{1}{2})^{2}}\Big[380\nonumber\\
&-3072\xi+12(1-q)(256\xi-43)+31(1-q)^{2}\nonumber\\
&-4(n+\tfrac{1}{2})^{2}(5q+17)(3-q))\Big]\Big\}+\mathcal{O}(l^{-3}).
\end{align}
To lowest order, we retrieve the result of Ref.~\cite{KonoplyaZhidenko} for minimal coupling but to next-to-leading order we have a coupling dependence. As in the case of the transmission coefficients we see that the quasi-normal mode dependence on the coupling constant is suppressed at large $l$. The latter limit further suggests the existence of long-lived modes in the limit $q\to 1$ since the large $l$ approximation of the damping can be made arbitrarily small by taking $q$ sufficiently close to unity.

A simple alternative approximation may be obtained by fitting the inverted potential to a potential whose bound states are known exactly, for example, the P\"oschl-Teller potential. This technique was developed by Ferrari and Mashhoon in order to estimate the quasi-normal modes of black holes \cite{FerrariMashhoon1, FerrariMashhoon2}. For Schwarzschild black holes, it gives better than a few percent accuracy compared with accurate numerical computations. If we denote
\begin{align}
\Lambda^{2}=l(l+1),\qquad \mu^{2}=2 q \xi+\tfrac{1}{2}(1-q),
\end{align}
then our potential may be written as
\begin{align}
V=\frac{\Lambda^{2}}{r^{2}}+\frac{\mu^{2}}{r^{2}}(\frac{b_{0}}{r}\Big)^{1-q}.
\end{align}
This potential remains invariant under the transformations
\begin{align}
\rho\to-i \,\rho,\quad \Lambda\to-i\,\Lambda,\quad b_{0}\to-i\,b_{0},\quad \mu\to-i\,\mu,
\end{align}
whence the wave equation (\ref{eq:wavepropeqn}) transforms to a bound-state problem with the inverted potential subject to vanishing boundary conditions at both spatial infinities. 
 \begin{figure}
\centering
\includegraphics[width=9.5cm]{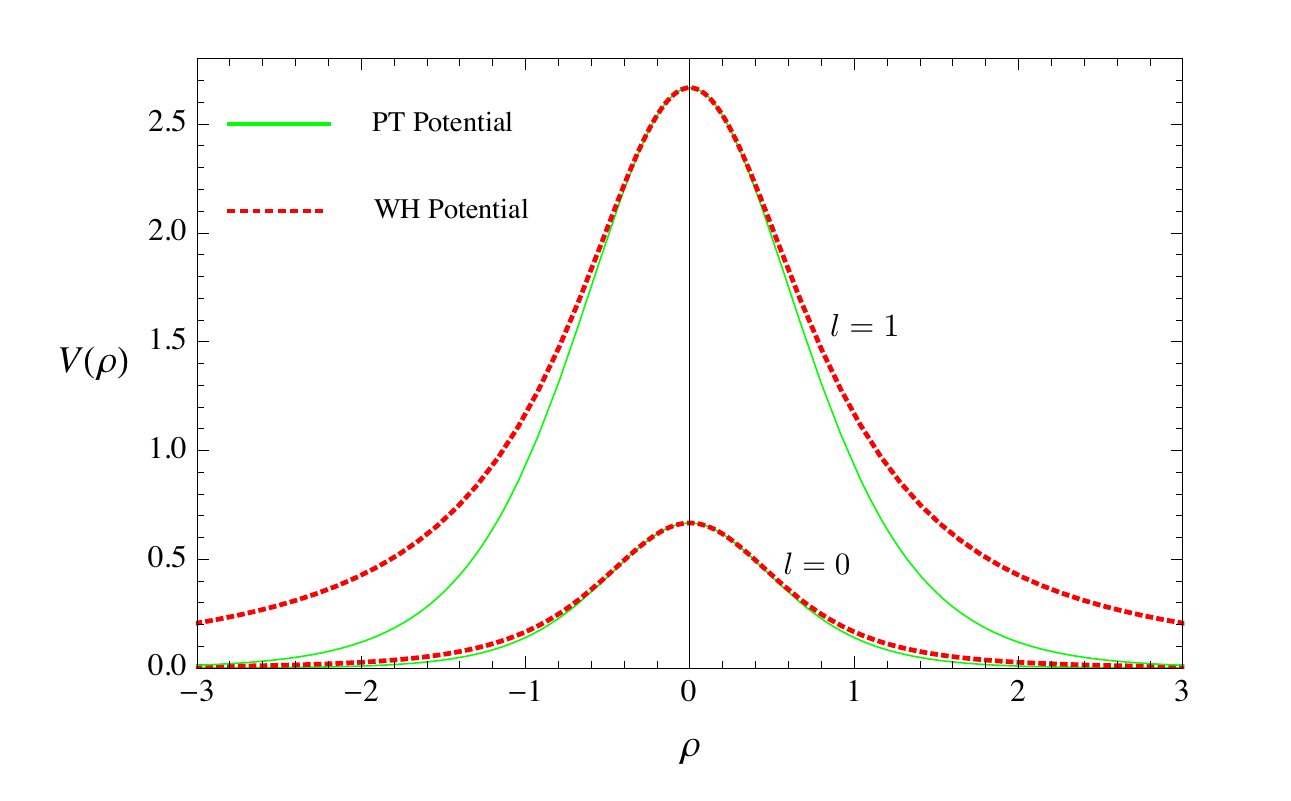}
\caption{{Plot of the wormhole potential (Red) and the P\"oschl-Teller (Green) potential for the modes $l=0$ and $l=1$. These plots are for the conformally coupled ($\xi=1/6$) scalar in the Ellis (q=-1) wormhole background. The throat radius has been set to unity.} }
\label{fig:ptpotential}
\end{figure} 

A good approximation to this potential is given by the P\"oschl-Teller potential (see Fig.~\ref{fig:ptpotential})
\begin{align}
V=\frac{V_{0}}{\cosh^{2}(\alpha \,\rho)},
\end{align}
with
\begin{align}
\label{eq:ptparam}
V_{0}&=\frac{\Lambda^{2}+\mu^{2}}{b_{0}^{2}},\nonumber\\
\alpha&=\frac{1}{2 b_{0}}\sqrt{\frac{(1-q)(2\Lambda^{2}-\mu^{2}(q-3))}{\Lambda^{2}+\mu^{2}}}.
\end{align}
The bound states of this potential are known \cite{PoschlTeller} to be
\begin{align}
\omega=-\alpha(n+\tfrac{1}{2})+(\tfrac{1}{4}\alpha^{2}+V_{0})^{1/2},
\end{align}
and hence analytically continuing back yields the following approximation for the quasi-normal modes
\begin{align}
\omega&=\Big(\frac{l(l+1)+2 q \xi+\tfrac{1}{2}(1-q)}{b_{0}^{2}}\nonumber\\
&-\frac{(1-q)(4l(l+1)-(q-3)(1-q+4q\xi))}{16 b_{0}^{2}(2l(l+1)+1-q+4q\xi)}\Big)^{1/2}\nonumber\\
&-i(n+\tfrac{1}{2})\Big(\frac{(1-q)(4l(l+1)-(q-3)(1-q+4q\xi))}{4 b_{0}^{2}(2l(l+1)+1-q+4q\xi)}\Big)^{1/2}.
\end{align}
This approximation fails to capture the overtone dependence of the oscillation frequency as does the analogous Schwarzschild approximation, but in the large $l$ limit gives the same leading order behavior as the WKB approximation.

We note that this procedure is straight-forward to adapt to compute the modes corresponding to a negative definite potential with a single minimum. For example, for the $l=0$ mode, this requires $4q\xi+1-q<0$. In this case, we require a set of transformations that reverse the sign of the potential, the appropriate choice is
\begin{align}
\rho\to-i\,\rho,\quad b_{0}\to-i\,b_{0}.
\end{align}
As before, this reduces the problem to finding the bound states of a negative definite potential subject to vanishing boundary conditions at both spatial infinities. Following the same procedure above, we obtain after analytically continuing back
\begin{align}
\omega=i\alpha(n+\tfrac{1}{2})+(-\tfrac{1}{4}\alpha^{2}-V_{0})^{1/2},
\end{align}
where $\alpha$ and $V_{0}$ are defined in Eq.~(\ref{eq:ptparam}) with $V_{0}<0$ in this case. It is clear that the $\textrm{Im}(\omega)>0$ and we have an unstable mode. This supports the analysis in Sec.~\ref{sec:stability} where we argued that Wald's stability criteria breaks down for certain values of the coupling constant. Of course, the above argument relies on an approximation to the quasi-normal modes, not the modes themselves. However, the fit of the wormhole potential to the P\"oschl-Teller potential is good even for the lowest lying modes (see Fig.~\ref{fig:ptpotential}) and we expect the approximation to be reasonably accurate, certainly accurate enough to determine the sign of the imaginary part.

\subsection{Static Self-Force in the Ellis Wormhole}
One interesting application of massless scalar fields propagating on wormhole background space-times is in the context of the self interaction of a point-like scalar particle of charge $Q$ and bare mass $m_{0}$ with its own scalar potential $\varphi(x)$. We assume that the mass of the particle is sufficiently small as to not perturb the wormhole space-time and hence we consider the background geometry fixed. If the scalar charge moves on a world-line $\gamma$ described by $z^{\mu}(\tau)$ where $\tau$ is proper time, then the action of this system is \cite{PoissonLR}
\begin{align}
S=S_{\textrm{field}}+S_{\textrm{particle}}+S_{\textrm{int}},
\end{align}
where
\begin{align}
S_{\textrm{field}}&=-\frac{1}{8\pi}\int (g^{\mu\nu}\nabla_{\mu}\varphi\nabla_{\nu}\varphi+\xi\,R\,\varphi^{2})\sqrt{-g}\,d^{4}x,\nonumber\\
S_{\textrm{particle}}&=-m_{0}\int_{\gamma}d\tau,\nonumber\\
S_{\textrm{int}}&=Q\int_{\gamma}\varphi(z(\tau))\,d\tau.
\end{align}
Stationarity of the action under variations of the field $\delta\varphi$ yields
\begin{align}
(\Box-\xi\,R)\varphi(x)=-\frac{4\pi\,Q}{\sqrt{-g}}\int_{\gamma}\delta^{4}(x-z(\tau))\,d\tau.
\end{align}
From the delta distribution source it is evident that the self-force, which is the gradient of the field evaluated at the location of the particle, is divergent and requires regularization. Regularization schemes typically involve subtracting a judiciously chosen parametrix from the gradient of the field before taking the limit as the field point approaches the world-line. The regularized field satisfies the homogeneous equation (\ref{eq:waveeqn}) corresponding to a massless scalar test field propagating on a fixed background which we have considered in detail in the previous sections.

Let us consider a particularly simple system consisting of a static scalar charge in the Ellis wormhole ($q=-1$) space-time with positive scalar coupling. The self-force has only a radial component which can be computed analytically \cite{Taylor} and is found to be
\begin{align}
\label{eq:fself}
 f=Q^{2}b_{0}\sqrt{2\xi}\cot(\sqrt{2\xi}\pi)\frac{\rho}{(\rho^{2}+b_{0}^{2})^{2}},
 \end{align}
 where $Q$ is the scalar charge and $\rho$ is the proper radial distance and $\xi\ge0$. The self-force exhibits some peculiar behavior depending on the value of the coupling constant. In particular, it is evidently singular whenever
 \begin{align}
 \xi=\frac{(n+1)^{2}}{2},\quad n\in \mathbb{Z},
 \end{align}
 and it vanishes whenever
 \begin{align}
 \xi=\frac{(2n+1)^{2}}{8},\quad n\in \mathbb{Z}.
 \end{align}
 However, we expect the self force to be a smooth function of the coupling since the wave equation is smooth in the coupling strength. On the other hand, the meaning of the spurious zeros of the self-force has remained unexplained, with the exception of the $\xi=1/8$ case for which we expect the static self-force to vanish since this is the value of conformal coupling in three dimensions and the static scalar charge satisfies a three-dimensional wave equation on a conformally flat space. (To see that the constant time slices of the Ellis wormhole are conformally spatially flat, make the coordinate transformation $\eta=\rho+\sqrt{\rho^{2}+a^{2}}$.)
 \begin{figure}
\centering
\includegraphics[width=8.5cm]{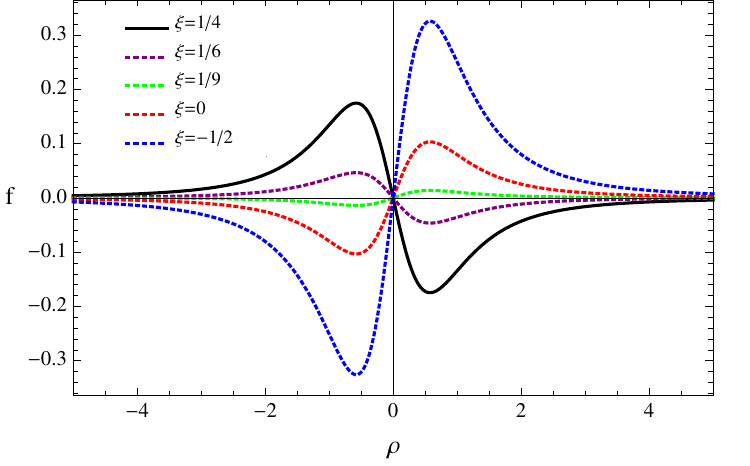}
\caption{{Plots of the static self-force in the Ellis wormhole space-time for various values of the coupling constant. The plot is in units where $b_{0}$ and the scalar charge $Q$ have been set to unity.} }
\label{fig:fself2D}
\end{figure}

\begin{figure}
\centering
\includegraphics[width=8.5cm]{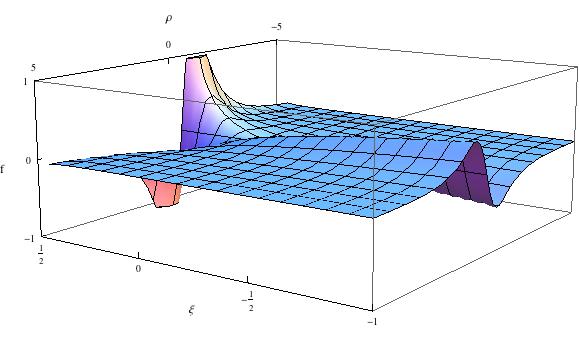}
\caption{{Plot of the self-force as a function of proper radius and coupling constant. It is a slowly varying function of $\xi$ except near the boundary of stability $\xi=1/2$ where its magnitude increases rapidly and without bound as $\xi\to1/2$.} }
\label{fig:fself3D}
\end{figure}

Since the potential generated by the scalar charge satisfies the wave equation (\ref{eq:waveeqn}), the stability analysis of Sec.~\ref{sec:stability} applies. Hence, we see that both the pathological and anomalous vanishing of the self-force lie outside the domain of stable solutions to this wave equation. If we restrict $\xi$ to the stable region $\xi<1/2$, then the self-force is a smooth function of the coupling and vanishes only at the three-dimensional conformal coupling value, as desired. Moreover, there is no lower bound on $\xi$ and we can analytically continue Eq.~(\ref{eq:fself}) for negative values of the coupling constant yielding
 \begin{align}
 \label{eq:fselfneg}
 f=Q^{2}b_{0}\sqrt{-2\xi}\coth(\sqrt{-2\xi}\pi)\frac{\rho}{(\rho^{2}+b_{0}^{2})^{2}},
 \end{align}
where we note the presence now of the hyperbolic cotangent. Combining Eqs.~(\ref{eq:fself}) and (\ref{eq:fselfneg}), it is clear that the self-force is a decreasing function of $\xi$ over the entire domain of stability. The force is attractive with respect to the throat in the narrow region $1/8<\xi<1/2$ and repulsive for all other values of the coupling $\xi<1/8$ (see Fig.~\ref{fig:fself2D}). From Fig.~\ref{fig:fself3D}, we see that, as a function of $\xi$, the magnitude of the force is slowly increasing as $\xi$ becomes more negative except on the domain of attraction $1/8<\xi<1/2$ where it increases rapidly.

Finally, we note that this anomalous coupling dependence of the static self-force in wormhole space-times was first realized by Bezerra and Khusnutdinov \cite{Khusnutdinov2} who considered a class of wormhole space-times which included the Ellis wormhole considered above. Analogous stability statements hold for the other wormholes that were considered in their paper and we claim that the self-force on a static scalar charge will not possess any poles within the domain of stability of massless scalar test fields on the background wormhole geometry. 

To illustrate this point, let us consider a particularly neat example where both the quasi-normal mode spectrum and the static self-force can be computed analytically, \textit{viz.}, a wormhole with an infinite short throat where the throat profile is given by 
\begin{align}
r(\rho)=|\rho|+b_{0},
\end{align}
where $b_{0}$ is the throat radius. It was shown in Ref.~\cite{Khusnutdinov2} that the static scalar self-force had poles for $\xi=(n+1)/4$. Since this is a particularly simple space-time (modulo the delta distribution curvature singularity at $\rho=0$), the mode functions in both the upper and lower regions are exactly soluble in terms of spherical Bessel functions. In order to compute the quasi-normal mode spectrum, one imposes purely outgoing radiation at both spatial infinities, i.e., $\chi_{\omega l}\sim e^{\pm i\omega \rho}$ as $\rho\to\pm\infty$. This condition results in the exact equation for the quasi-normal modes
\begin{align}
\label{eq:qnmexact}
(l+1-4\xi)H^{(1)}_{l+1/2}(\omega b_{0})-\omega b_{0}H^{(1)}_{l-1/2}(\omega b_{0})=0
\end{align}
where $H^{(1)}_{\nu}(z)$ is the Hankel function of the first kind. The quasi-normal mode spectrum of this particular wormhole is peculiar in that there are a finite number of frequencies for each mode. This is in contrast to black holes and smooth wormhole space-times where there are an infinite set of frequencies for each $l$-mode. In fact it is straight-forward to show that there are $l+1$ frequencies for each $l$-mode. Since the Hankel functions of half-integer order can be written explicitly as a finite sum (see \cite{GradRiz} for example), we can recast Eq.~(\ref{eq:qnmexact}) as
\begin{align}
(l+1-4\xi)\sum_{k=0}^{l}i^{k-l-1}\frac{c_{k,l}}{(b_{0}\omega)^{k+1}}-\sum_{k=0}^{l-1}i^{k-l}\frac{c_{k,l-1}}{(b_{0}\omega)^{k}}=0
\end{align}
where
\begin{align}
c_{k,l}=\begin{cases}
\displaystyle{\frac{(l+k)!}{2^{k}k!(l-k)!}} & \quad k=0,1,2,...,l\nonumber\\
0 & \quad k=l+1,l+2,...
\end{cases}
\end{align}
This amounts to solving a polynomial of degree $l+1$ and hence there are $l+1$ solutions. 

Again in order to constrain $\xi$ so that solutions of the wave equation remain bounded for all time, we need only consider the $l=0$ mode, for which there is only one pure imaginary frequency
\begin{align}
\omega=\frac{i}{b_{0}}(4\xi-1)
\end{align}
which corresponds to pure exponential damping for $\xi<1/4$ and to exponential growth for $\xi>1/4$. Hence, as in the case of the Ellis wormhole, the lowest lying pole of the static self-force coincides with the boundary of stability of the solutions to the wave equation. Restricting to $\xi<1/4$ cures the pathological behavior of the static self-force.

\section{Conclusions}
In this paper, we have solved the causal geodesic equations in a class of ultra static wormhole space-times defined by the profile functions $b(r)=b_{0}^{1-q}/r^{q}$, where $b_{0}$ is the throat radius and $q$ is the so-called shape exponent. In terms of these geometric parameters and initial conditions for the position and direction of the geodesic, we have characterized the fate of the geodesic, i.e., whether it is reflected, trapped on an unstable orbit on the wormhole throat or propagates through the wormhole to the other universe. We found that the space of initial directions that traverse the wormhole decreases with decreasing shape exponent, i.e., as the magnitude of the curvature of the throat increases, so too does the angular momentum potential barrier and hence reflection of the geodesic becomes increasingly favorable. We have described how to visualize these geodesics in an embedding space and have given a sample of embedding plots for various wormhole parameters and initial conditions.

In the second part of this article, we studied test scalar fields arbitrarily coupled to this class of background wormhole space-times. We have shown that the wave equation for massless scalar test fields admits stable solutions only when the coupling constant and shape exponent satisfy $2 q \xi+\tfrac{1}{2}(1-q)>0$. Restricted to the domain of stability, we computed transmission coefficients and quasi-normal modes using the fourth-order WKB method. In particular, we focused on the dependence of these quantities on the coupling strength which we found to be suppressed in the geometric optics limit. Finally, we revisited the calculation of the static scalar self-force on the Ellis wormhole space-time ($q=-1$) which was known to display some anomalous dependence on the coupling strength, in particular, the self-force vanishes for $\xi=(2n+1)^{2}/8$ and diverges for $\xi=(n+1)^{2}/2$ where $n=0,\pm 1,\pm2,...$ We showed that this was a consequence of na\"ively computing the self-force for values of the coupling constant outside the domain of stability. When restricted to $\xi<1/2$, the force is everywhere regular and vanishes only at the three-dimensional conformal value as expected. Moreover, we argued that this was a generic feature of ultra-static wormhole space-times and showed by explicit computation of exact quasi-normal modes in a wormhole with an infinitely short throat that the lowest lying pole of the static self-force is equivalent to the boundary of the stable solutions.
\\

\acknowledgments
I would like to thank \'Eanna Flanagan for his careful reading of the manuscript and for his helpful suggestions. This work has also benefited from conversations with fellow colleagues at Cornell, in particular, I would like to thank David Nichols, Leo Stein and Barry Wardell.

This work is supported by the Irish Research Council under the ELEVATE scheme which is co-funded by the European Commission under the Marie Curie Actions program. I would also like to acknowledge support from NSF grant PHY-1068541.

\bibliographystyle{apsrev4-1}
\bibliography{database}
\end{document}